\documentclass[12pt,thmsa]{article}
\usepackage{amsmath}
\usepackage{graphicx}
\usepackage{amsfonts}
\usepackage{amssymb}

\begin{document}

\title{Objections to the Unified Approach to the Computation of Classical Confidence Limits}
\author{G\"{u}nter Zech\thanks{E-mail:zech@physik.uni-siegen.de}\\Universit\"{a}t Siegen, D-57068 Siegen, Germany}
\maketitle
\begin{abstract}
Conventional classical confidence intervals in specific cases are unphysical.
A solution to this problem has recently been published by Feldman and
Cousins\cite{feld98}. We show that there are cases where the new approach is
not applicable and that it does not remove the basic deficiencies of classical
confidence limits.
\end{abstract}

\section{ Introduction}

Feldman and Cousins propose a new approach to the computation of
\emph{classical confidence bounds} which avoids the occurrence of unphysical
confidence regions, one of the most problematic features of the conventional
classical confidence limits. In addition it unifies the two procedures
``\emph{computation of confidence intervals}'' and ``\emph{computation of
confidence limits}''. The unified treatment represents a considerable
improvement compared to the conventional classical method and has already been
adopted by several experiments and is recommended by the Particle Data Group
\cite{pdg98}. However, it has serious deficiencies.

\section{Basic idea of the unified approach}

We consider the example of section B of Ref. \cite{feld98}. For a Gaussian
resolution function $P(x;\mu)$ we define for each mean $\mu$ an interval
$x_{1}<x<x_{2}$ with the property
\begin{equation}
\int_{x1}^{x2}P(x;\mu)dx=\alpha
\end{equation}
where $\alpha$ is the confidence level. For a measurement $\hat{x}$ all values
$\mu$ with the property $x_{1}(\mu)<\hat{x}<x_{2}(\mu)$ form the confidence
interval. The intervals have the property that the true values are covered in
the fraction $\alpha$ of a large number of experiments. The freedom in the
choice of the interval inherent in the relation (1) is used to avoid
unphysical limits. (Usually the limits $x_{1},x_{2}$ are fixed by choosing
central intervals.) In case that only one limit can be placed inside the
allowed parameter space, upper (or lower) limits are computed. The data and
the selected level $\alpha$ unambiguously fix the bounds and whether bounds or
limits are given. The probability bounds are defined by an ordering scheme
based on the likelihood ratio. In the case of discrete parameters an analogous
procedure is applied with some additional plausible conventions. The complete
recipe is too complicated to be discussed in a few words. The reader has to
consult the original publication.

\section{Objections to the unified approach}

The new approach has very attractive properties, however, there are also
severe limitations most of which are intrinsic in the philosophy of classical statistics.

\subsection{Inversion of significance}

In some cases less significant data can provide more stringent limits than
more informative data.

As an example we present in the following table the $90\%$ confidence upper
limits for a Poisson distributed signal from data with no event found ($n=0$)
for different background expectations of mean $b$.

The experimental information on the signal $s$ is the same in all four cases
independent of the background expectation since no background is present. For
the case $n=0,b=3$ the unified approach avoids the unphysical negative limit
of the conventional classical method but finds a limit which is more
significant than that of an experiment with no background expected and twice
the flux.

If in the $n=0,b=3$ experiment by an improved analysis the background
expectation is reduced, the limit becomes worse.

The reason for this unsatisfactory behavior is related to the violation of the
likelihood principle\footnote{A detailed discussion of the likelihood
principle and references can be found in \cite{berg84} and \cite{basu88}.} by
the classical methods. All four cases presented in the table have common
likelihood functions $L\sim e^{-s}$ of the unknown signal up to an irrelevant
multiplicative constant depending on $b$.\begin{table}[ptb]%
\begin{tabular}
[c]{|l|l|l|l|l|}\hline
& n=0, b=0 & n=0, b=1 & n=0, b=2 & n=0, b=3\\\hline
standard classical & \multicolumn{1}{|c|}{2.30} & \multicolumn{1}{|c|}{1.30} &
\multicolumn{1}{|c|}{0.30} & \multicolumn{1}{|c|}{-0.70}\\\hline
unified classical & \multicolumn{1}{|c|}{2.44} & \multicolumn{1}{|c|}{1.61} &
\multicolumn{1}{|c|}{1.26} & \multicolumn{1}{|c|}{1.08}\\\hline
uniform Bayesian & \multicolumn{1}{|c|}{2.30} & \multicolumn{1}{|c|}{2.30} &
\multicolumn{1}{|c|}{2.30} & \multicolumn{1}{|c|}{2.30}\\\hline
\end{tabular}
\caption{Confidence limits for Poisson distributed data with $n$ observed
events and expected background with mean $b$. }%
\end{table}

\subsection{Difficulties with two-sided bounds}

Let us assume a measurement $\hat{x}=0$ of a parameter $x$ with a physical
bound $-1<x<1$ and a Gaussian resolution of $\sigma=1.1$. (This could be for
example a track measurement by a combination of a proportional wire chamber
and a position detector with Gaussian resolution.) The unified approach fails
to give 68.3\% confidence bounds or limits.

\subsection{Difficulties with certain probability distributions}

The prescription for the definition of the probability intervals may lead to
disconnected interval pieces. A simple example for such a distribution is the
superposition of a narrow and a wide Gaussian
\[
P(x;\mu)=\frac1{\sqrt{2\pi}}\left\{  0.9\exp\left(  -(x-\mu)^{2}/2\right)
+\exp\left(  -(x-\mu)^{2}/0.02\right)  \right\}
\]
with the additional requirement of positive parameter values $\mu$. It will
produce quite odd confidence intervals.

Another simple example is the linear distribution
\[
P(x;\theta)=\frac12(1+\theta x)
\]
where the parameter $\theta$ and the variate $x$ are bound by $|\theta|\leq1$
and $|x|\leq1$. (The variable $x$ could be the cosine of a polar angle.)
Values of $\theta$ outside its allowed range produce negative probabilities.
Thus the likelihood ratio which is used as a ordering scheme for the choice of
the probability interval is undefined for $|\theta|>1$. Remark that also the
conventional classical confidence scheme fails in this case.

Similarly all digital measurements like track measurements with proportional
wire chambers or TDC time registration cannot be treated. Since the
probability distributions are delta-functions the bounds are undefined.

\subsection{Restriction due to unification}

Let us assume that in a search for a Susy particle a positive result is found
which however is compatible with background within two standard deviations.
Certainly one would prefer to publish an upper limit to a measurement contrary
to the prescription of the unified method.

\subsection{Difficulty to use the error bounds}

Errors associated to a measurement usually are used to combine results from
different experiments or to compute other parameters depending on them. There
is no prescription how this can be done in the unified approach. Averaging of
data will certainly be difficult due to the bias introduced by asymmetric
probability contours used to avoid unphysical bounds. Feldman and Cousins
propose to use the conventional classical limits for averaging. Thus two sets
of errors have to be documented.

\subsection{Restriction to continuous variables}

It is not possible to associate a classical confidence to discrete hypothesis.

\subsection{Subjectivity}

The nice property of a well defined coverage depends on pre-experimental
analysis criteria: The choice of the selection criteria and of the confidence
level as well as the decision to publish have to be done independently of the
result of the experiment. This requirement is rather naive.

\section{Conclusions}

There are additional difficulties to those discussed above: The elimination of
nuisance parameters and the treatment of upper Poisson limits with uncertainty
in the background predictions pose problems. These may be tractable but
certainly introduce further complications. The computation of the limits will
be very computer time consuming in most cases. The essential objections,
however, are those mentioned in sections 3.1, 3.3 and 3.5. It is absolutely
intolerable that significant limits can be obtained with poor data and it is
also essential to have useful error intervals. Feldman and Cousins are aware
of the difficulties related to the inversion of significance and to biased
errors and propose to publish additional information. This certainly is a
sensible advice but does not justify classical limits. Most of the
deficiencies of the conventional classical method remain unresolved in the
unified approach.

The experimental information relative to a parameter can be documented by its
likelihood function. The log-likelihood functions of different experiments can
easily be combined without introducing biases simply by adding them. In most
cases the likelihood function can be parametrized in a sensible way, as is
common practice, by the parameters which maximize the likelihood and the
values at $1/\sqrt{e}$ of the maximum. The latter define an error interval. In
the case of Poisson limits the Bayesian limits with constant prior (see Table
1) provide a useful parametrization which avoids the difficulties of section
3.1. These pragmatic procedures, however, do not allow to associate a certain
coverage to the intervals or limits. Coverage is the magic objective of
classical confidence bounds. It is an attractive property from a purely
esthetic point of view but it is not obvious how to make use of this concept.

.

\end{document}